\documentclass[webconf, onecolumn]{svjour}
\usepackage{graphics}
\usepackage[varg]{txfonts} 
\usepackage[latin1]{inputenc}
\session-title{AO4ELT}
\begin{document}
\title{Extreme Adaptive Optics in the mid-IR: The METIS AO system}
\author{
R. Stuik\inst{1}\fnmsep\thanks{\email{stuik@strw.leidenuniv.nl}} \and 
L. Jolissaint\inst{1} \and 
S. Kendrew\inst{1} \and 
S. Hippler\inst{2} \and 
B. Brandl\inst{1} \and 
L. Venema\inst{3} \and 
the METIS team
} 
\institute{
Leiden Observatory, Leiden University, P.O. Box 9513, 2300 RA Leiden, The Netherlands \and
Max Planck Institute for Astronomy, K\"onigstuhl 17, 69117 Heidelberg, Germany \and
ASTRON, Postbus 2, 7990 AA Dwingeloo, the Netherlands
} 
\abstract{
Adaptive Optics at mid-IR wavelengths has long been seen as either not necessary or easy. The impact of atmospheric turbulence on the performance of 8-10 meter class telescopes in the mid-IR is relatively small compared to other performance issues like sky background and telescope emission. Using a relatively low order AO system, Strehl Ratios of larger than 95\% have been reported on 6-8 meter class telescopes. Going to 30-42 meter class telescopes changes this picture dramatically. High Strehl Ratios require what is currently considered a high-order AO system. Furthermore, even with a  moderate AO system, first order simulations show that the performance of such a system drops significantly when not taking into account refractivity effects and atmospheric composition variations. Reaching Strehl Ratios of over 90\% at L, M and N band will require special considerations and will impact the system design and control scheme of AO systems for mid-IR on ELTs. In this paper we present an overview of the effects that impact the performance of an AO system at mid-IR wavelengths on an ELT and simulations on the performance and we will present a first order system concept of such an AO system for METIS, the mid-IR instrument for the E-ELT.
} 
\maketitle
\section{Introduction}
\label{sec:intro}
METIS, the Mid-infrared ELT Imager and Spectrograph \cite{brandl08}, is currently in its phase A study as one of the candidate first-light instruments for the European Extremely Large Telescope (E-ELT) \cite{gilmozzi07}. METIS will feature several observational modes. It will provide diffraction limited imaging over a field of 18$\times$18'' for L, M, and N band, which includes a Four Quadrant Phase Mask Coronagraph (4QZOG), a low-resolution ($R \approx 5000$) long-slit spectrometer and a polarimeter for the N-band. Furthermore METIS will contain a high-resolution Integral Field Unit (IFU) ($\geq 0.4\times1.6''$) spectrograph for the L and M-band ($2.9 - 5.3\, {\mu}$m, $R \approx 100,000$), while optionally a high-resolution N-band IFU spectrograph is being considered. Using the 42-meter aperture of the E-ELT, minimal warm surfaces and using the M4 and M5 mirror of the E-ELT for a--relatively--high-order Adaptive Optics (AO) system, METIS in its current design gives a similar sensitivity to Spitzer in imaging and low-resolution spectroscopy and with its high-resolution spectrograph will provide unprecedented line sensitivity. METIS will provide complementarity to JWST and Spitzer. The design of METIS is optimized for both galactic science cases ({\it e.g.} conditions in the early solar system, formation and evolution of proto-planetary disks and properties of exo-planets) and extragalactic science cases ({\it e.g.} the growth of super-massive black holes).

\section{METIS AO}
\label{sec:metisao}
\subsection{Requirements}
\label{subsec:requirements}
The current generation of 8-10 meter class telescopes achieve diffraction limited performance without an AO system, while with an optimized AO system, routinely Strehl Ratios over 95\% in the mid-IR range are reported \cite{biller05,skemer09}. With the 4-5 times larger aperture on the E-ELT, METIS {\it does} require a high-order AO system to meet its scientific goals. Apart from providing a good correction under varying seeing conditions, stabilizing the PSF in time, it will also provide a correction for wind shake of the telescope. 

Two solutions are currently foreseen for the METIS AO system, both utilizing the internal deformable mirror (M4) and image stabilizing mirror (M5) of the E-ELT. As a base line, METIS will be fitted with an internal WFS, providing an Single Conjugate AO (SCAO) system. A separate study is currently ongoing for an external Laser Tomography AO system, which can provide extended sky coverage and more uniform correction in the field. The METIS AO system will need to overcome several challenges: 
\begin{itemize}
\item It will need to provide broad-band correction in the range between 3 and 13\,$\mu$m.
\item It will need to provide a correction ranging from a Strehl Ratio of over 80\% at L-Band and over 93\% at 10 $\mu$m for standard seeing condition of 0.8''.
\item It will need to provide an very stable image position for coronagraphy (better than 5 mas RMS image stability).
\item It will need to provide a high sky coverage, to be provided by excellent sensitivity of the wavefront sensor (WFS) and potentially usage of the LTAO mode.
\item It will provide an unobstructed science field by both providing a 'cold' dichroic for splitting off the light for the WFS as well as maintaining the Laser Guide Star pick-up outside the METIS scientific field of view.
\end{itemize}

\subsection{Basic AO Performance}
\label{subsec:basicperformance}
\begin{figure}[t]
\begin{center}
\resizebox{0.75\columnwidth}{!}{%
\includegraphics{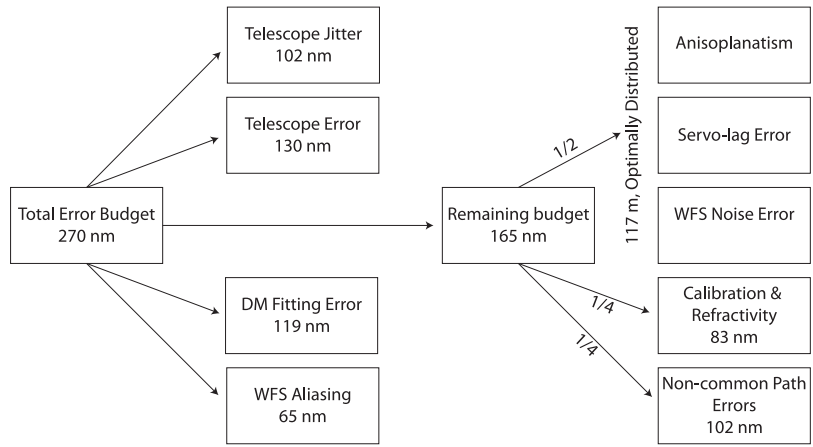} }
\end{center}
\caption{First order error budget for the METIS AO system. The total error budget is given by the required L-band performance, while the DM and telescope errors are mainly given by the latest design of the E-ELT and internal deformable mirrors. The remaining error budget is used to optimize the system for guide star magnitude and speed of the system.}
\label{fig:0}       
\end{figure}
\begin{figure}[htbc]
\begin{center}
$\begin{array}{cc@{\hspace{0.5cm}}c}
\resizebox{0.45\columnwidth}{!}{%
\includegraphics{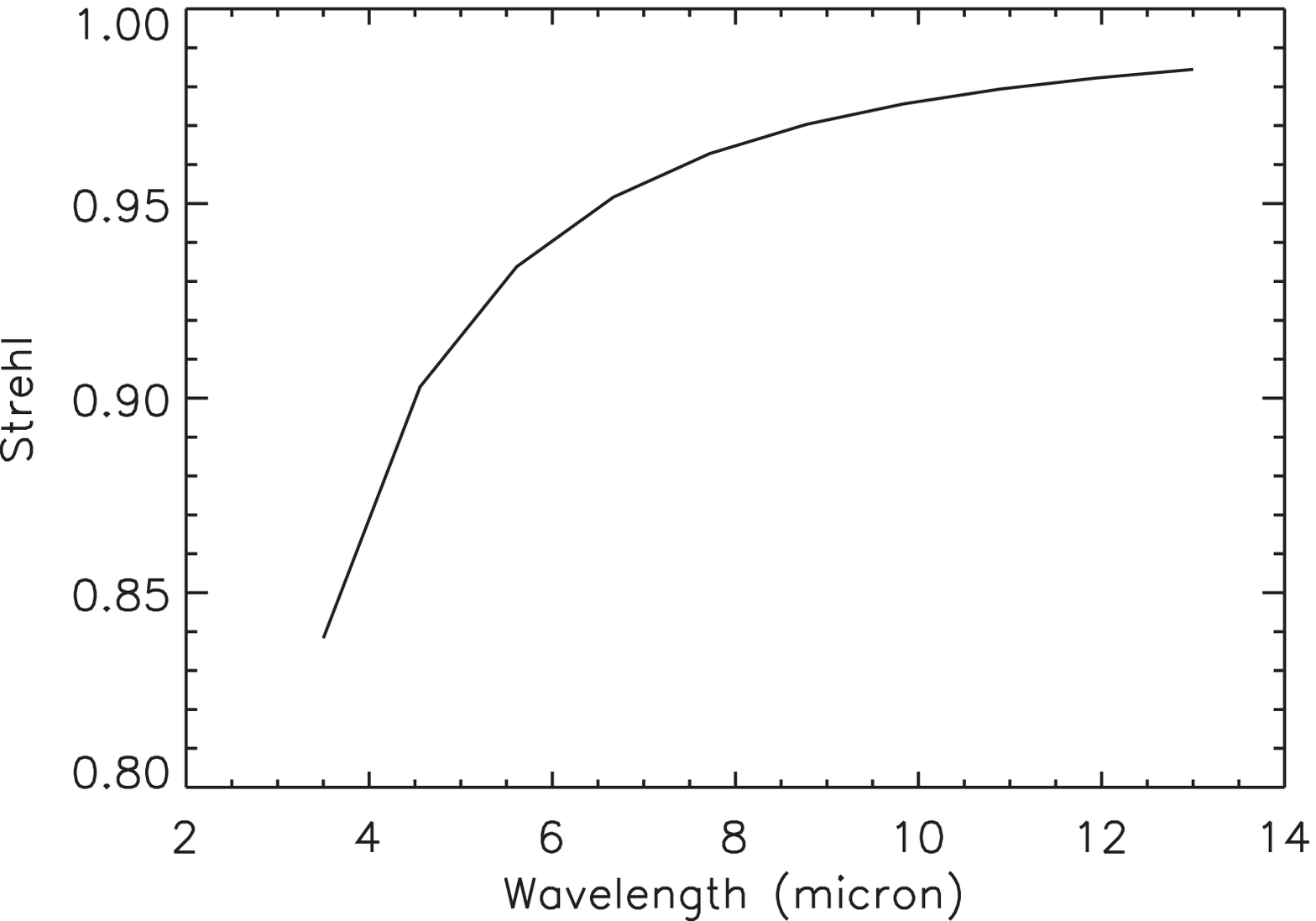}} & &
\resizebox{0.45\columnwidth}{!}{%
\includegraphics{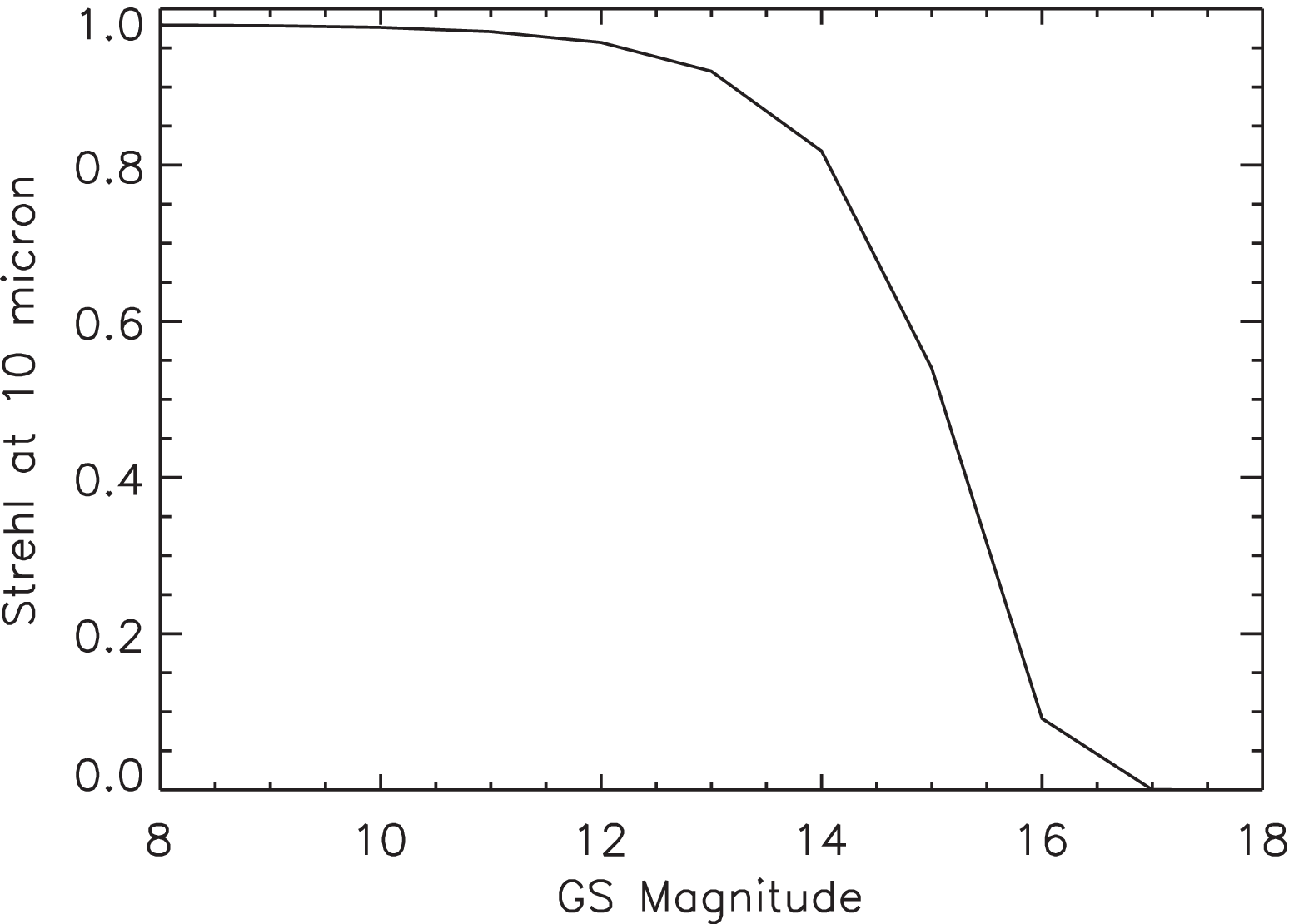}} \\ 
\multicolumn{1}{r}{(a)}& & \multicolumn{1}{r}{(b)}\vspace{0.5cm}\\
\resizebox{0.45\columnwidth}{!}{%
\includegraphics{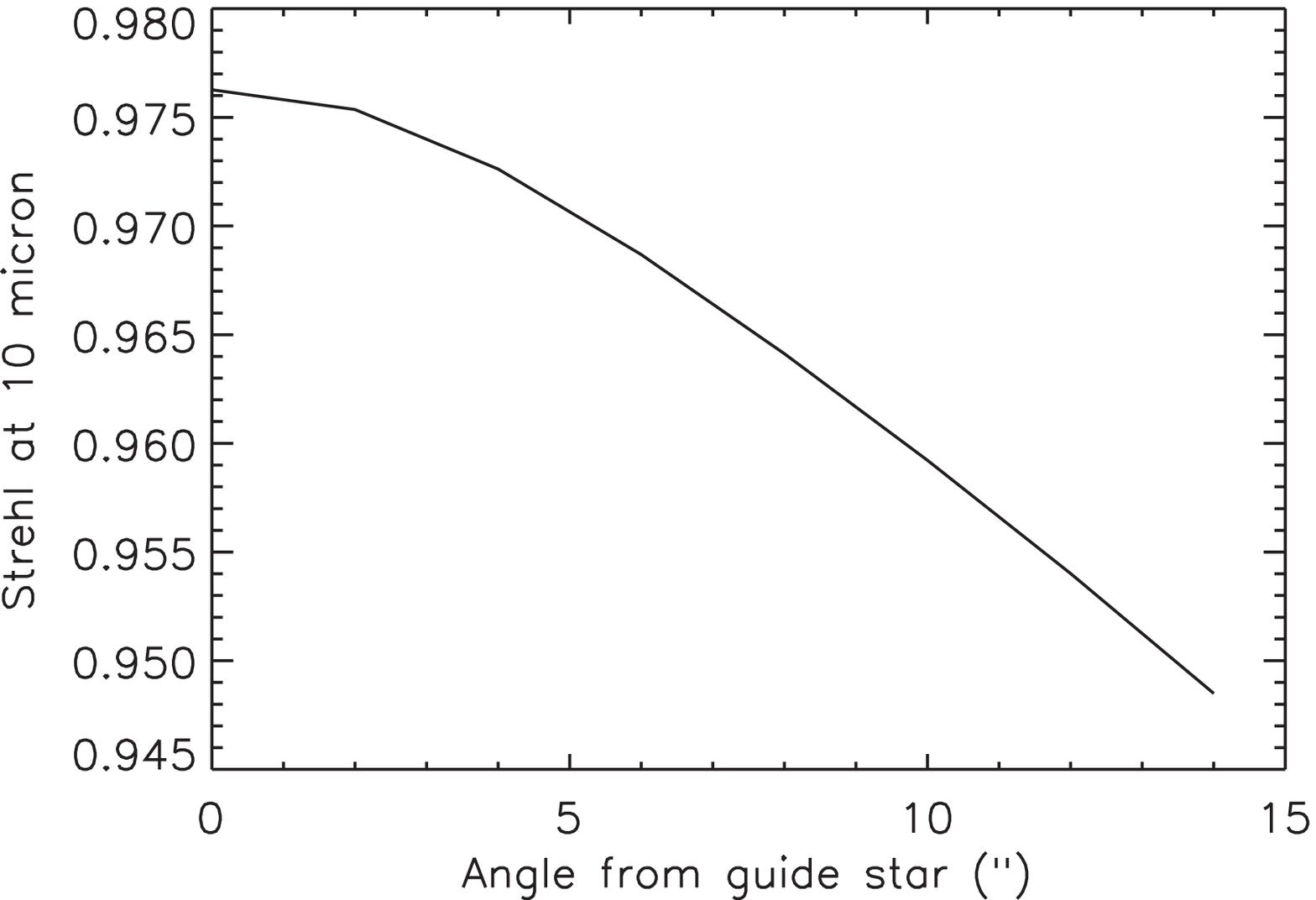}} & & \\
\multicolumn{1}{r}{(c)}& & 
\end{array}$
\end{center}
\caption{Simulated performance of the METIS AO system. Panel {\it(a)} shows the on-axis performance of METIS as a function of wavelength assuming a bright visible Natural Guide Star. Panel {\it(b)} shows the on-axis performance of METIS at a wavelength of 10\,$\mu$m as a function of the guide star magnitude. Panel {\it(c)} shows the performance in the field at a wavelength of 10\,$\mu$m, assuming a bright on-axis guide star. All plots ignore the effects of fluctuations in the index of refraction.}
\label{fig:1} 
\end{figure}

The basic AO performance of METIS was estimated using PAOLA \cite{jolissaint06}, an analytical simulation code based on the spatial power spectrum approach. The performance of METIS was computed in its SCAO configuration and included a full error budget, but excluded effects of water vapor. Under these conditions METIS meets its main requirements; Using a bright visible guide star (m$_{\rm V} < 13$), a loop frequency of $>$ 100\,Hz and a sub-sampling of the WFS matched to the actuators of the internal DMs of the E-ELT, the METIS AO system is expected to achieve an on-axis Strehl Ratio between 83 (L-band) and 97\% (N-band) and less than 5\% variation in Strehl Ratio in the field. Figure \ref{fig:1} shows the performance plots for this basic METIS system. Panel {\it(a)} shows the simulated Strehl Ratio as a function of wavelength for the center of the field of view, assuming a central Natural Guide Star. It shows the typical behavior of a strong decrease of the performance towards shorter wavelengths. Panel {\it(b)} shows the decrease in performance of the METIS AO system as a function of the guide star magnitude at a wavelength of 10\,$\mu$m, in the center of the field of view. The performance of METIS is maintained up to a guide star magnitude of (m$_{\rm V} = 13$) after which the performance drops strongly. The third panel, {\it(c)}, shows the performance over the full field of METIS at a wavelength of 10\,$\mu$m, again assuming a central bright guide star. The performance drops slightly to the edge of the field of view, but remains within the requirements.

\subsection{Impact of Water Vapor}
\label{subsec:watervapour}
New in the METIS AO system, compared to existing AO systems, is the impact of composition turbulence. This comes mainly in the form of (fast) variations in the Precipitable Water Vapor (PWV) content, both because the wavelengths for which correction is required ({\it i.e.} the mid-IR are strongly influenced by water vapor themselves) as well as the increased size of the telescope with respect to the current generation, which amplifies effects currently not or barely seen at 8-10 meter class telescopes. 

\begin{figure}[htbc]
\begin{center}
$\begin{array}{c@{\hspace{0.5cm}}c}
\resizebox{0.45\columnwidth}{!}{%
\includegraphics{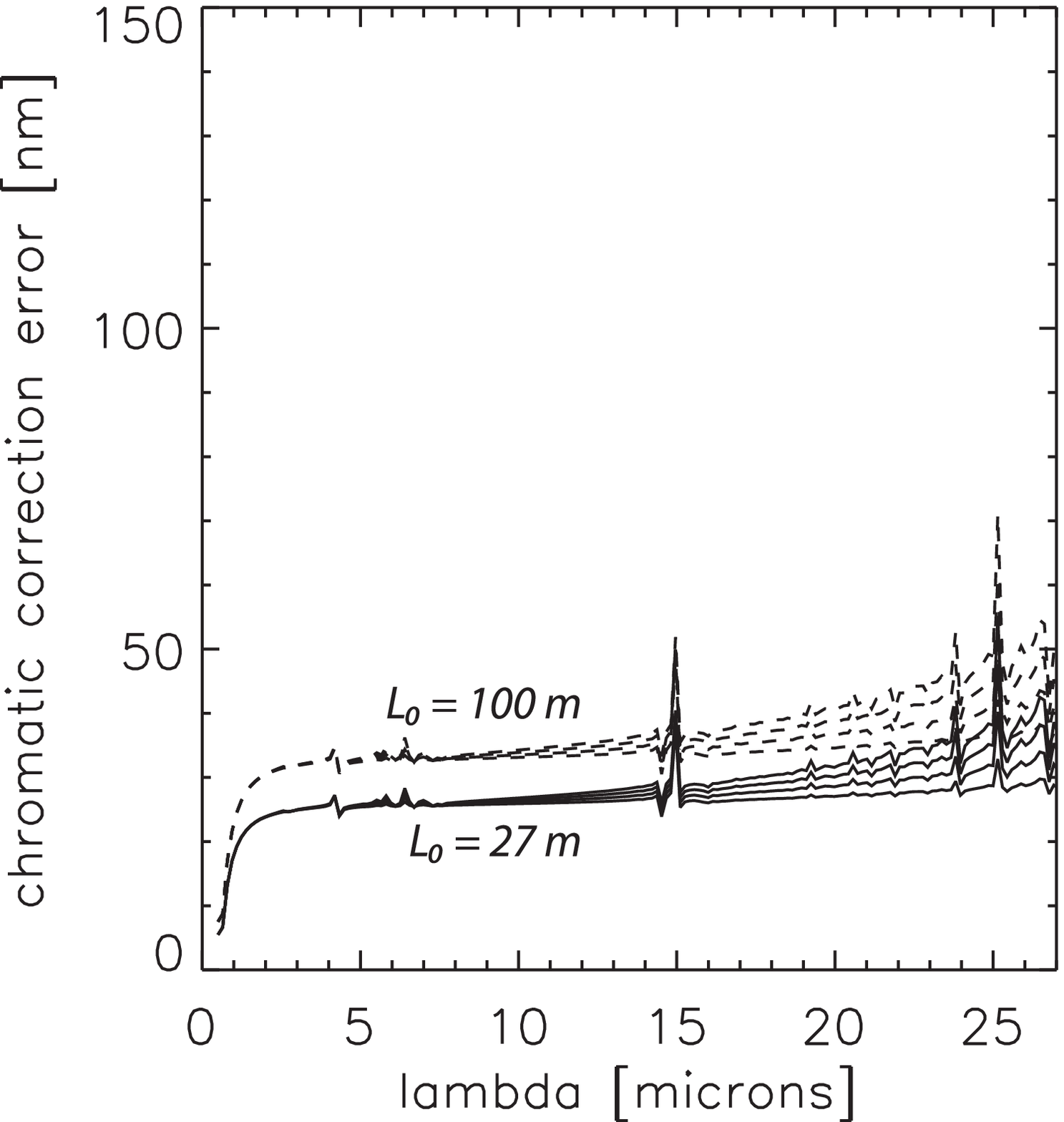}} &
\resizebox{0.45\columnwidth}{!}{%
\includegraphics{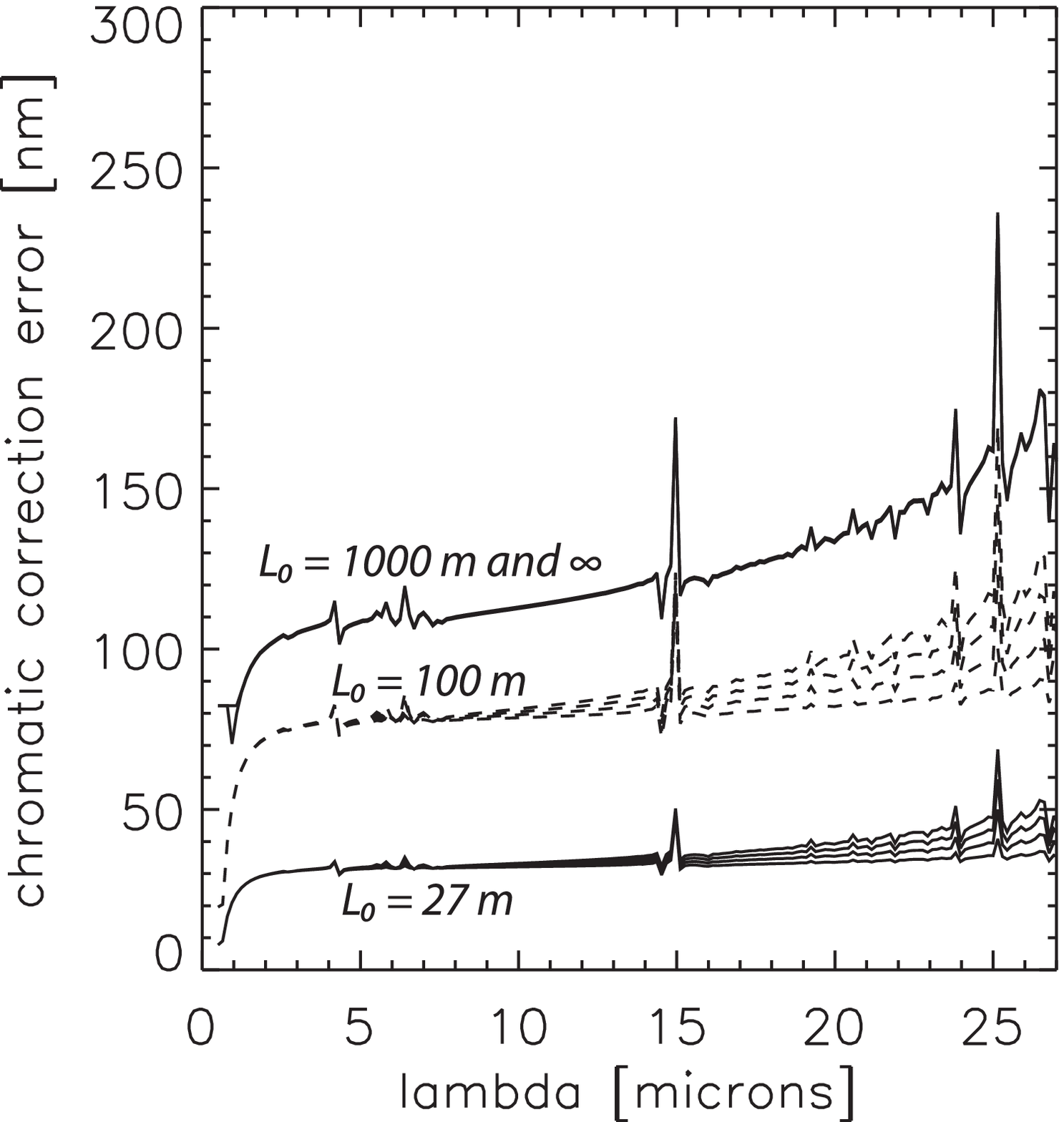}} 
\end{array}$
\end{center}
\caption{Chromatic correction error at mid-IR wavelengths, when using a visible light wavefront sensor and a seeing of 0.65'', for a 6.5 meter telescope (left panel) and a 42 meter telescope (right panel) for various values of the outer scale and various values of the humidity. For each of the traces for different outer scale values, the relative humidities are (from bottom to top) 5, 10, 15 and 20\%.}
\label{fig:2}       
\end{figure}

Water vapor and water vapor fluctuations impact the performance of METIS in several ways. Standard atmospheric dispersion, the effect that light at different wavelengths is bent over a slightly different angle due to index of refraction variations as a function of wavelength, is an effect that is already taken into account in existing instruments and can be relatively easy corrected using an Atmospheric Dispersion Corrector (ADC). This correction is required for both the WFS, where the extension of the source due to atmospheric dispersion needs to be corrected to maintain the WFS sensitivity, as well as for the relative displacement between the WFS wavelength and the science wavelength. Variations in the water vapor composition will have an impact on the atmospheric dispersions, although only the larger but slower components will give a noticeable effect; the change in atmospheric dispersion is currently estimated at $\leq 10$ mas, which is small, but can, especially for coronagraphy, not be neglected. A second effect is the chromatic optical path difference (OPD) error, which is caused by changes in the index of refraction along the path through the atmosphere. Contrary to the standard atmospheric dispersion, the OPD errors are even present at zenith. The impact is the worst when using visible wavelengths for wavefront sensing for correcting at mid-IR wavelengths, but the impact is reduced by the outer scale. The resulting rms wavefront errors are given in Figure \ref{fig:2}. As can be seen, the resulting errors do not exceed 120 nm RMS within the wavelength band of METIS and the resulting Strehl Ratio multiplier at 10\,$\mu$m is 0.9996 at a 'typical' value of the outer scale of 27 m, see also \cite{jolissaint09}. 

\begin{figure}[htbc]
\begin{center}
\resizebox{0.60\columnwidth}{!}{%
\includegraphics{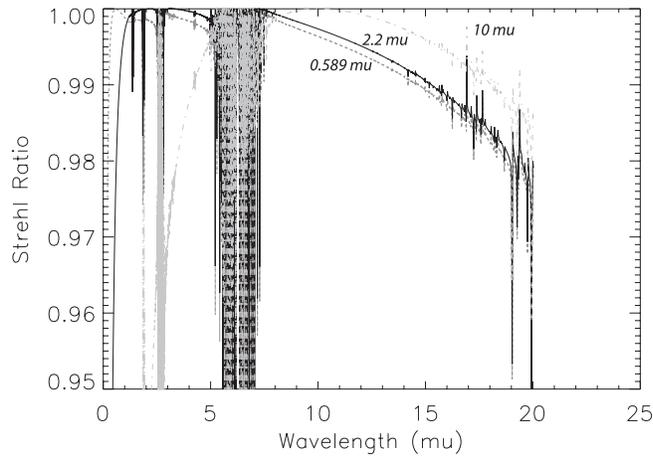} }
\end{center}
\caption{Strehl Ratio multiplier due to the impact of water vapor turbulence as a function of wavelength for three different wavefront sensing wavelengths, namely $0.589\,\mu$m (dark grey, dashed line), $2.2\,\mu$m (black, solid line) and $10\,\mu$m (light grey, dot-dashed line). Note that the METIS wavelength range is $3 - 13\,\mu$m.} 
\label{fig:3}       
\end{figure}

Chromatic anisoplanatism is the effect that the light travels through a slightly different light path through the atmosphere. This effect, which is again aggravated by changes in the composition, can be--at least partly--corrected, as discussed in \cite{devaney08}. The last effect is composition turbulence, mainly caused by fast fluctuations in the water vapor content. The current estimates are based on measurements for ALMA \cite{nikolic08} and radiometer measurements by \cite{colavita04} and range from $11 - 22\,\mu$m RMS PWV fluctuations for a 42 meter telescope. Figure \ref{fig:3} shows the resulting Strehl Ration multiplier due to water vapor turbulence at a strength of $20\,\mu$m RMS PWV fluctuations for three different wavefront sensing wavelengths. Using a sensing wavelength in either the visible or the infrared, a decrease in Strehl Ratio by a factor of 0.92 can be expected. The effect can be reduced by sensing at the science wavelength, as demonstrated by the curve for $10\,\mu$m. Note that the performance at the shortest science wavelengths of the METIS, near $3\,\mu$m, the impact of wavefront sensing at $10\,\mu$m is significantly stronger and would decrease the usefulness of simultaneous observations over the full wavelength range of METIS. Currently there is no solution to compensate for this effect over broader wavelength bands, even when sensing at the science wavelength. Composition turbulence is still under investigation, but most importantly will require significantly more measurement data on the distribution and magnitude of PWV fluctuations.

\subsection{METIS Baseline AO System}
\label{subsec:baseline}
The current baseline for the METIS AO system is a visible light Shack-Hartmann WFS, in order to use the synergies with other WFSs being developed for the E-ELT. Furthermore, we are investigating to also include a infrared WFS to enhance the sky coverage for deeply embedded sources ({\it i.e.} (V-K) $>$ 5) and using the internal AO system for METIS as WFS for and external LTAO system. Currently the AO system will contain a pupil-derotator and an ADC for both correcting the wavelength range of the WFS as well as stabilizing the offset between WFS and science channel. The WFS elements will be matched to the actuators of the deformable mirror in the E-ELT. METIS will not use a field selector, while it will use the M4 and M5 mirrors of the E-ELT for both image correction as well as image stabilization. No facilities are currently foreseen to correct for composition fluctuations. The main goal is to provide a simple and stable AO system that will provide excellent and reliable performance for METIS.

\acknowledgement
\label{sec:acknowledgements}
We gratefully acknowledge funding from NOVA, the Netherlands Research School for Astronomy. We are also grateful to the METIS consortium for supporting and participating in this work.

\end{document}